\documentclass[aps,prl,twocolumn,showpacs,superscriptaddress,reprint]{revtex4-1}
\usepackage{graphicx}
\usepackage{natbib}
\usepackage{amsmath}
\usepackage{amssymb}
\usepackage{bm}
\usepackage{color}

\begin{document}

\title{Revealing multiple gaps in the electronic structure of USb$_2$ using femtosecond optical pulses}

\author{J. Qi} 
\affiliation{Los Alamos National Laboratory, Los Alamos, NM 87545, USA}
\author{T. Durakiewicz} 
\affiliation{Los Alamos National Laboratory, Los Alamos, NM 87545, USA}
\author{S. A. Trugman} 
\affiliation{Los Alamos National Laboratory, Los Alamos, NM 87545, USA}
\author{J.-X. Zhu} 
\affiliation{Los Alamos National Laboratory, Los Alamos, NM 87545, USA}
\author{P. S. Riseborough} 
\affiliation{Temple University, Philadelphia, PA 19121, USA}
\author{R. Baumbach} 
\affiliation{Los Alamos National Laboratory, Los Alamos, NM 87545, USA}
\author{E. D. Bauer} 
\affiliation{Los Alamos National Laboratory, Los Alamos, NM 87545, USA}
\author{K. Gofryk} 
\affiliation{Los Alamos National Laboratory, Los Alamos, NM 87545, USA}
\author{J.-Q. Meng} 
\affiliation{Los Alamos National Laboratory, Los Alamos, NM 87545, USA}
\author{J. J. Joyce} 
\affiliation{Los Alamos National Laboratory, Los Alamos, NM 87545, USA}
\author{A. J. Taylor} 
\affiliation{Los Alamos National Laboratory, Los Alamos, NM 87545, USA}
\author{R. P. Prasankumar} 
\affiliation{Los Alamos National Laboratory, Los Alamos, NM 87545, USA}

\date{\today}

\begin{abstract}
Ultrafast optical spectroscopy is used to study the antiferromagnetic $f$-electron system USb$_2$. We observe the opening of two charge gaps at low temperatures ($\lesssim 45$ K), arising from renormalization of the electronic structure. Analysis of our data indicates that one gap is due to hybridization between localized $f$-electron and conduction electron bands, while band renormalization involving magnons leads to the emergence of the second gap. These experiments thus enable us to shed light on the complex electronic structure emerging at the Fermi surface in $f$-electron systems.
\end{abstract}

\maketitle
Exotic phenomena, such as unconventional superconductivity, the heavy fermion state, or the elusive "hidden order" phase of URu$_2$Si$_2$, can emerge from many-body interactions in $f$-electron systems \cite{Mathur_Nat_1998, Chandra_Nat_2002, Curro_Nat_2005, Yang_Nat_2008}. These phenomena, governed by strong electronic correlation and complex interactions between the electronic and bosonic degrees of freedom, are often associated with the dual nature (localized vs. itinerant) of the $f$-electrons. The itinerant response is usually related to the nature of the Fermi surface, where very small band renormalization effects, captured theoretically through the electron self-energy, often escape experimental detection due to lack of resolution. However, these minute changes in the Fermi surface cannot be ignored, as they can dramatically modify the properties of $f$-electron materials, leading to the emergence of hidden order parameters, increase of the effective mass, superconductivity, and the appearance of the heavy fermion state \cite{HOref, Basov, Choi}. A deeper understanding of these phenomena thus depends on a detailed knowledge of the electronic structure near the Fermi surface in $f$-electron systems.

Antiferromagnetic USb$_2$ \cite{Trzebiatowski_BAPSSSC_1963, Henkie_APP_1972} is an excellent candidate for exploring these issues, as it is a moderately correlated electron system with a quasi-2D electronic structure that exhibits characteristics of both localized and itinerant 5$f$ electrons \cite{Wawryk_PhiMag_2006, Aoki_JPSJ_1999, Aoki_PhiMag_2000, Aoki_PhysB_2000, Wisniewski_JPSJ_2001, Durakiewicz_EPL_2008}. Moreover, previous angle-resolved photoemission spectroscopy (ARPES) studies indicated that bosons could play an important role in band renormalization \cite{Durakiewicz_EPL_2008, Durakiewicz_PRB_2004, Yang_PhiMag_2009}. These studies produced the first measurements of the self-energy in 5$f$ electron systems and a model for boson-mediated band renormalization. However, they could only explore the region near the center of the Brillouin zone ($k=0$) and could not specify the nature of the boson involved. This directly motivated our search for new physical properties (e.g., the opening of a gap or changes in the quasiparticle (QP) effective mass) arising from boson-mediated many-body interactions, in order to understand the complex 5$f$ band structure in USb$_2$ over all $k$-space.

Ultrafast optical spectroscopy (UOS) has been quite successful in providing such information, offering insight into the physics of different strongly correlated materials \cite{Basov_RMP_2011}, such as superconductors (SCs) \cite{Han_PRL_1990, Kabanov_PRB_1999, Chia_PRL_2007} and heavy fermions (HFs) \cite{Demsar_JPC_2006, Talbayev_PRL_2010}.  In particular, by measuring the temperature ($T$)-dependent QP dynamics and analyzing the data with the Rothwarf-Taylor (RT) model \cite{Rothwarf_PRL_1967, Kabanov_PRL_2006}, one can accurately extract small changes in the electronic structure (e.g., the opening of a small gap in the DOS), even away from the Brillouin zone center.

Here, we present the first ultrafast time-resolved differential reflectivity $\Delta R(t)/R$ measurements on USb$_2$, taken from room temperature down to 6 K. We reveal multiple gaps opening in the density of states (DOS), associated with emergent QP states at/near the Fermi level. We also observe coherent phonon oscillations, previously unobserved in metallic uranium systems, the analysis of which not only illustrates the temperature evolution of the gap structures, but also demonstrates that \textit{magnons} can be involved in band renormalization. 

In our experiment, high quality USb$_2$ samples, with N\'{e}el temperatures of $T_N\sim$ 203 K, were grown from Sb flux and cleaved along the $ab$ plane prior to measurement. The transient photoinduced reflectivity change, $\Delta R/R$, was measured using a Ti:sapphire laser oscillator that produced $\sim$55 femtosecond (fs) pulses at a center wavelength of 830 nm ($\sim$1.5 eV) and a repetition rate of 80 MHz. The pump and probe beams were cross-polarized, with fluences of 0.3 $\mu$J/cm$^2$ ($\sim$5x10$^{-5}$ photoexcited carriers/unit cell) and 0.02 $\mu$J/cm$^2$, respectively. The pump beam was incident at a $\sim$10 degree angle to the sample normal, with the probe beam directed along the normal.

\begin{figure}[h]
\includegraphics[width=7cm]{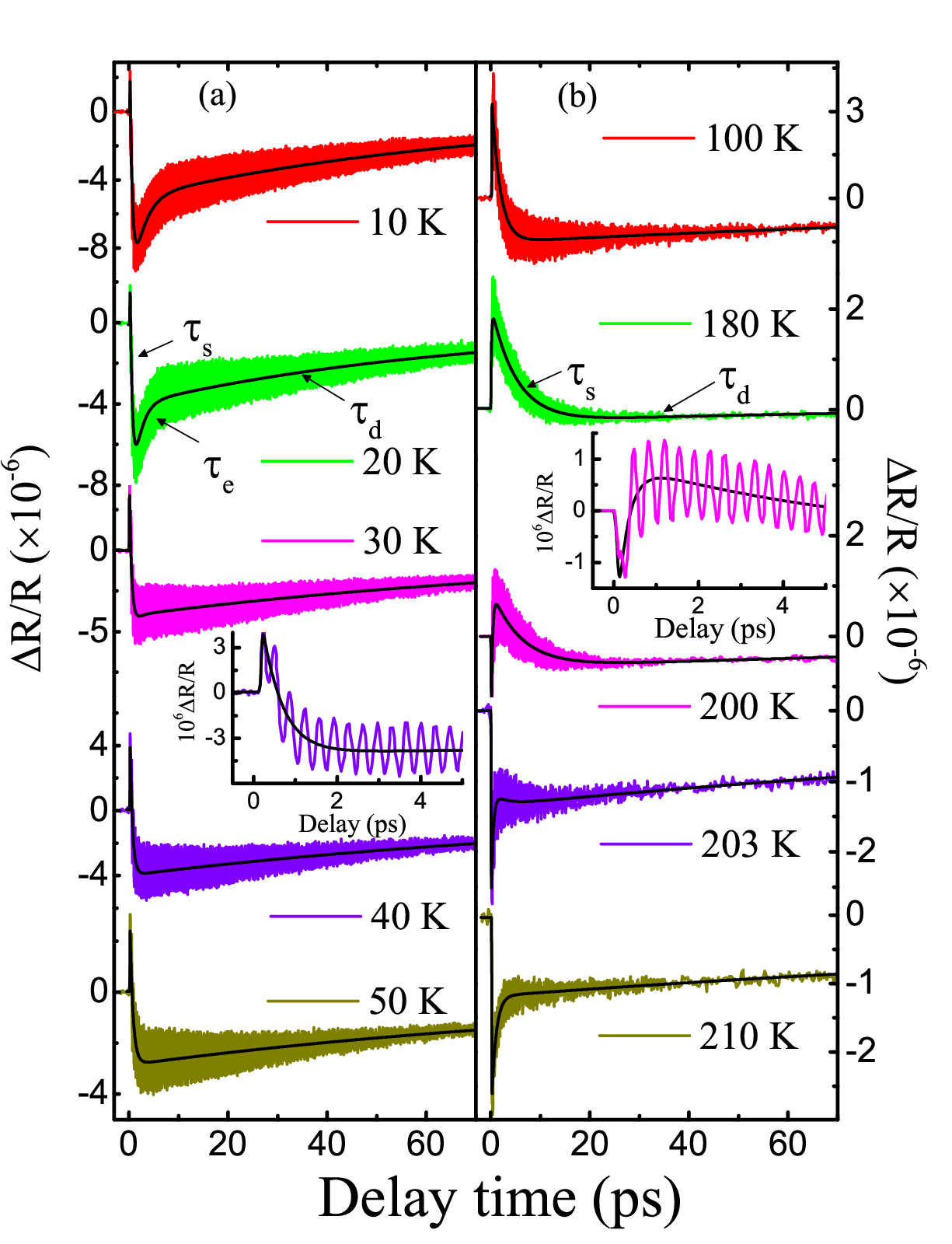}
\caption{\label{fig:deltaR} Temperature-dependent $\Delta R/R$ data for USb$_2$. The solid lines show the extracted non-oscillatory background decay. The insets show the dynamics at short timescales for two temperatures (40 K and 200 K). The arrows indicate the corresponding decay processes.}
\vspace*{-0.4cm}
\end{figure}

Figure \ref{fig:deltaR} shows the measured $\Delta R/R$ signals at (a) low and (b) high temperatures. Upon photoexcitation, the $\Delta R/R$ signal changes nearly instantaneously due to a rise in the temperature of the Fermi surface, as excited carriers rapidly equilibrate via electron-electron scattering \cite{Demsar_JPC_2006}. This initial change reverses sign when the temperature crosses $\sim$200 K ($\sim T_N$). This is consistent with the reconstruction of the Fermi surface at $T_N$, as previously seen in dHvA measurements \cite{Aoki_PhiMag_2000, Aoki_JPSJ_1999}. At longer delays, after equilibration of the photoexcited carriers, the $\Delta R/R$ signals exhibit a damped ultrafast oscillation superimposed on a non-oscillating background decay.

We will first focus on the non-oscillatory relaxation for $T\lesssim T_N$, since it does not significantly change in the paramagnetic state ($T\gtrsim T_N$). Below $\sim T_N$, this relaxation can be fitted with three exponential decays convoluted with a Gaussian laser pulse $G(t)$ (FHWM$\simeq$55 fs): $\Delta R/R=[A_se^{-t/\tau_s}+A_ee^{-t/\tau_e}+A_de^{-t/\tau_d}+A_0]\otimes G(t)$, ($A_s>0$, $A_e<0$, $A_d<0$). Here, $A_e$ and $\tau_e$ represent a relaxation process that only appears below a critical temperature of $T^\dagger\sim$32 K, and $\tau_s$ and $\tau_d$ are the time constants of the initial fast decay and very slow relaxation, respectively (see Fig. \ref{fig:deltaR}). 

\begin{figure}
\includegraphics[width=7cm]{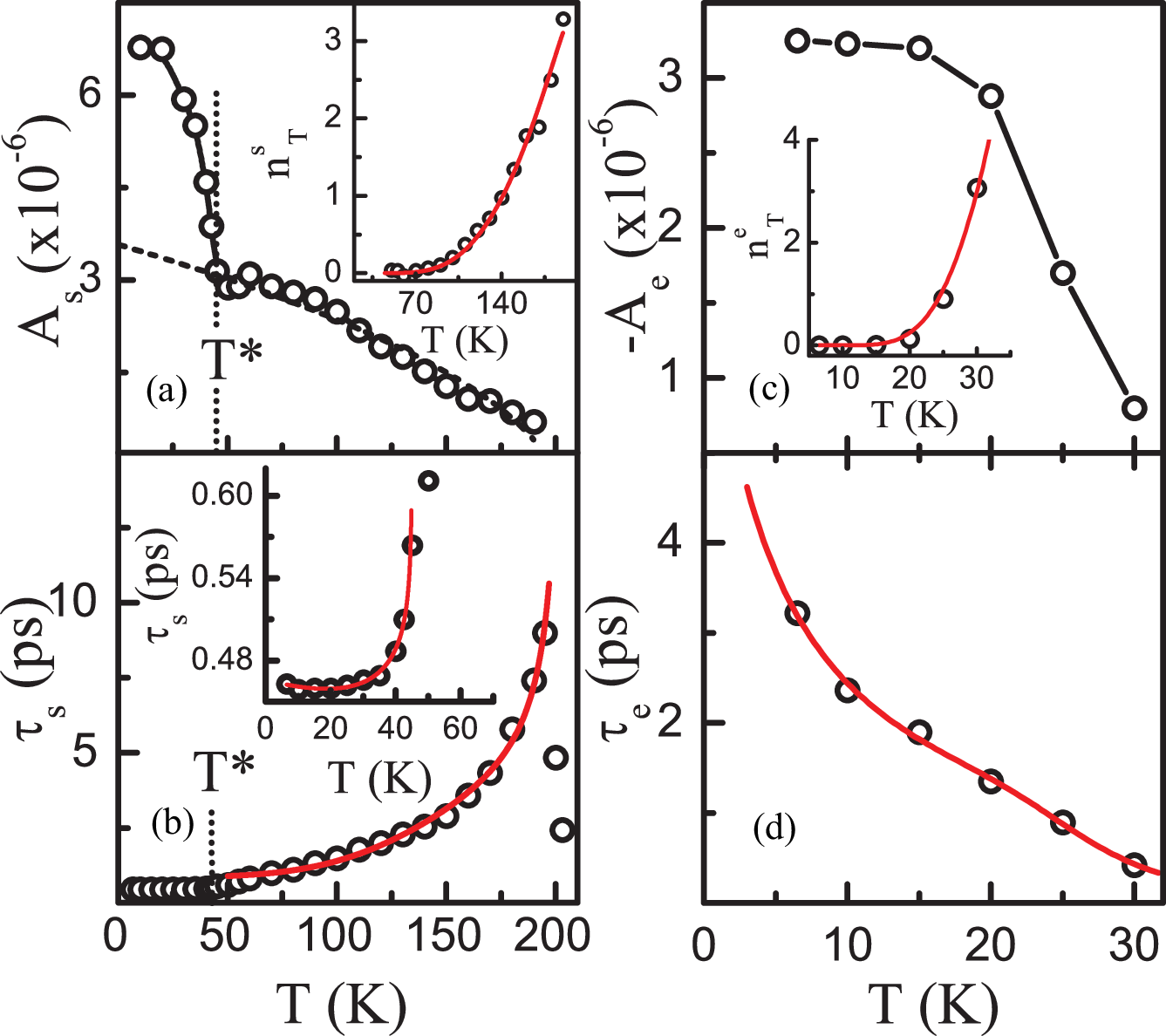}
\caption{\label{fig:RTfit} $T$-dependence of amplitudes $A_j$ and excited quasiparticle densities $n_T^j$ ($j=s,e$), corresponding to the relaxation times $\tau_s$ ((a) and (b)) and $\tau_e$ ((c) and (d)), respectively. The red solid lines are best fits to the data using the RT model. The dashed line in (a) is an amplitude fit using a BCS-like $T$-dependence to obtain the value of $A_s(T=0)$ for fitting $\tau_s(T>T^*)$ \cite{Chia_PRL_2007}.}
\vspace*{-0.2cm}
\end{figure}

We find that the relaxation characterized by $\tau_d$ has a timescale of at least a few hundred picoseconds (ps) at all temperatures, which is likely due to thermal diffusion, as in similar measurements on other strongly correlated systems \cite{Demsar_JPC_2006, Qi_APL_2012}. Here, we will focus on the other two decay processes that occur on shorter timescales ($t\lesssim$10 ps). We clearly observe from Figure \ref{fig:RTfit} that: (a) $\tau_s$ increases continuously with $T$, and shows a sharp upturn at $T_N$ (nearly diverging there); (b) As $T$ decreases through a temperature $T^*$ ($\simeq45$ K), $A_s$ strongly increases, while $\tau_s$ decreases significantly (inset to Fig. 2(b)); (c) $A_e$ and $\tau_e$, which first appear below $T^\dagger$, increase rapidly as $T$ decreases. Similar temperature-dependent behaviour has been observed in superconductors \cite{Chia_PRL_2007, Kabanov_PRB_1999} and HF compounds \cite{Demsar_JPC_2006, Chia_PRB_2006}, and was attributed to the opening of a gap ($\Delta$) in the DOS. However, the detailed physical mechanism behind the gap opening varies for different systems and thus demands careful analysis, as described below.

Quasiparticle relaxation in a system with a narrow gap, such as superconductors and HFs, can be approximated by the Rothwarf-Taylor model \cite{Rothwarf_PRL_1967}. Here, the recovery process is governed by the decay of electrons with energies larger than the gap, via the emission of high  frequency bosons (HFBs) that can subsequently re-excite electron-hole pairs (EHPs). The RT model has been applied to many materials \cite{Chia_PRL_2007, Demsar_JPC_2006, Kabanov_PRL_2006}, using the equations
\begin{align}
n_T(T)=A(0)/A(T)-1,
\\
\tau^{-1}(T)\propto[\delta(\beta n_T+1)^{-1}+2n_T](\Delta+\alpha T\Delta^4)
\label{eq:RT-model},
\end{align}
where $n_T(T)$ is the density of thermally excited quasiparticles and $\alpha$, $\beta$ and $\delta$ are fitting parameters. In these equations, we employ the standard form of $n_T$: $n_T\propto(T\Delta)^pe^{-\Delta/T}$ \cite{Demsar_JPC_2006}, where the choice of $p$ ($0<p<1$) depends on the shape of the DOS.   

We can use the RT model to gain more insight into the processes characterized by $\tau_s$ and $\tau_e$. To model both of these processes, we use $p=0.5$, which represents the DOS with a shape similar to that of BCS superconductors, as previously used for heavy fermion systems (Fig. \ref{fig:RTfit}) \cite{Demsar_JPC_2006}. We then fit $\tau_e$ (Fig. \ref{fig:RTfit}(c,d)) using a $T$-independent constant gap $\Delta_e$, with a value of 11.8 meV. Modeling $\tau_s$ is more complicated, as the sharp change across $T^*$ necessitates that we fit the data differently above and below $T^*$. At higher temperatures ($T>T^*$), we can fit $\tau_s$ using a BCS-like $T$-dependence for the gap: $\Delta_s\simeq 46.1(1-T/T_N)^{0.5}$ (meV). However, at lower temperatures $(T\lesssim T^*)$, $\tau_s$ can be fit with neither a BCS-like $T$-dependent gap nor a constant gap. Therefore, we assume a gap with a simple $T$-dependent form: $\Delta_s^*=\Delta_s^*(0)(1-T/T^*)^\eta$, where $\eta$ is also a fitting parameter. With this assumption, we find that $\tau_s(T\lesssim T^*)$ can be reproduced well using $\Delta_s^*\simeq 12.8(1-T/T^*)^{0.05}$ (meV). The excellent agreement between the experimental results and the curve fits confirms our initial expectation of a gap opening.

The quasi-divergence of $\tau_s$ at the N\'{e}el temperature suggests a quasiparticle gap opening (measured by $\Delta_s$) due to the onset of magnetic order. In USb$_2$, this magnetic order contains contributions from both spin and orbital polarization. Simultaneously, conditions for a possible Fermi surface nesting appear because of parallel boundaries existing between hole and electron pockets in this system \cite{Aoki_JPSJ_1999, Aoki_PhiMag_2000, Aoki_PhysB_2000, Lebegue_PRB_2006}. This is similar to the previous observation of a spin density wave gap in the itinerant antiferromagnetic actinide UNiGa$_5$ \cite{Chia_PRB_2006}. 

\begin{figure}[t!]
\includegraphics[width=7cm,height=8cm]{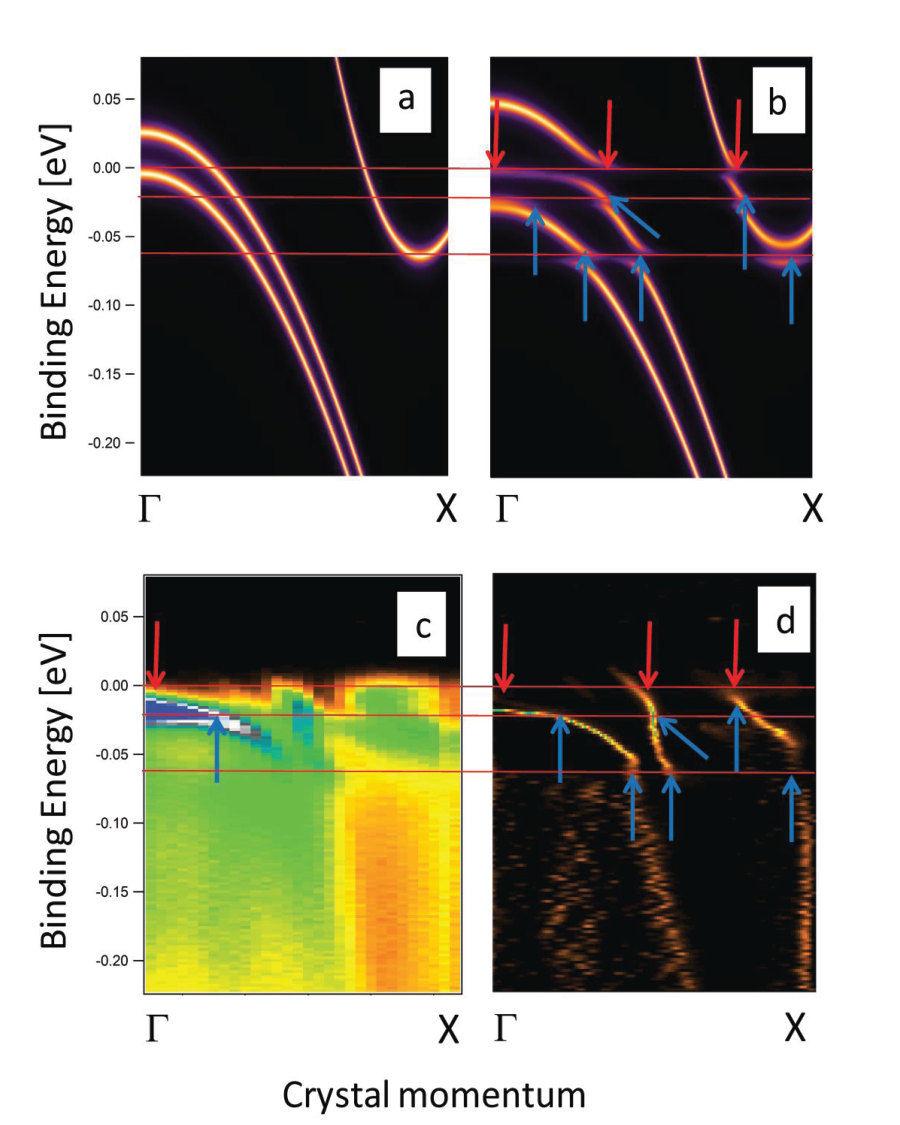}
\caption{\label{fig:BandStructure} Multiple gaps and kinks revealed by ARPES measurements on USb$_2$ at 12 K. Panels (a) and (b) show the calculated bare bands and the renormalized bands when including boson-mediated band renormalization \cite{Threebands}, respectively. Panel (c) shows the original ARPES data \cite{Durakiewicz_EPL_2008, Yang_PhiMag_2009}, including the single gap-and-kink structure. Panel (d) shows data from (c) reduced with the 2D curvature method \cite{Zhang_RSI_2011}. Multiple gap structures are marked with red arrows, with their energy scales indicated by red horizontal lines. Blue arrows mark kinks in the dispersion due to band renormalization.}
\vspace*{-0.4cm}
\end{figure}

More interestingly, our analysis indicates that two additional gaps open up at lower temperatures ($\Delta_s^*$ and $\Delta_e$), which we will focus on below. Previous ultrafast optical studies on HFs \cite{Demsar_JPC_2006} have shown that such small gaps are typically due to hybridization between the localized 5$f$ electrons and the conduction electrons. However, ARPES studies on USb$_2$ \cite{Durakiewicz_EPL_2008, Durakiewicz_PRB_2004, Yang_PhiMag_2009} have revealed an extremely narrow band below the Fermi level and a kink-like feature in the band dispersion near the zone center (Fig.~\ref{fig:BandStructure}(c)) that cannot be explained by such hybridization alone. Ref.~\cite{Yang_PhiMag_2009} shows that electron-boson mediated processes also contribute to these effects, where the bosons participate in interband electron scattering. Therefore, we need to consider both of these potential contributions in discussing the origin of $\Delta_s^*$ and $\Delta_e$.

We have performed a theoretical analysis of the electronic structure that incorporates both hybridization between $f$ electron bands and conduction bands as well as boson-mediated processes. To illustrate this, the calculated results along the $\Gamma$-X direction are shown in Figures \ref{fig:BandStructure}(a) and (b). In our calculations, the non-interacting or bare bands are obtained from local density approximation (LDA) calculations, and boson-mediated band hybridization is introduced via interband scattering \cite{Yang_PhiMag_2009}. Fig. \ref{fig:BandStructure} clearly shows that boson-mediated band renormalization can lead to multiple gaps and/or kink-like structures away from zone-center at/near the Fermi level. More specifically, our calculations produce indirect charge gaps at the Fermi level with a magnitude of $\sim$10-15 meV for both hole- and electron-like bands. 

These predicted values agree well with current experimental findings. Previously, it was not possible to compare the calculation to the original  ARPES data over the full Brillouin zone \cite{Durakiewicz_EPL_2008, Durakiewicz_PRB_2004, Yang_PhiMag_2009}, since the features away from zone center were very hard to discern due to the rapidly decreasing signal intensity for high $k$ values. Here, we use a recently introduced data reduction method involving the 2D Fermi surface curvature \cite{Zhang_RSI_2011} to identify those features (Fig.~\ref{fig:BandStructure}(d)), allowing us to show that the complex multi-gap structure predicted by theory indeed agrees well with the ARPES data. Thus, we propose that boson-mediated many-body interactions play a prominent role in band renormalization at/near the Fermi surface, and contribute to the low-$T$ gap openings observed here. However, understanding the nature of the QP states associated with the band gaps $\Delta_s^*$ and $\Delta_e$, as well as the type of boson involved, requires further evaluation.

\begin{figure}[h]
\includegraphics[width=7cm]{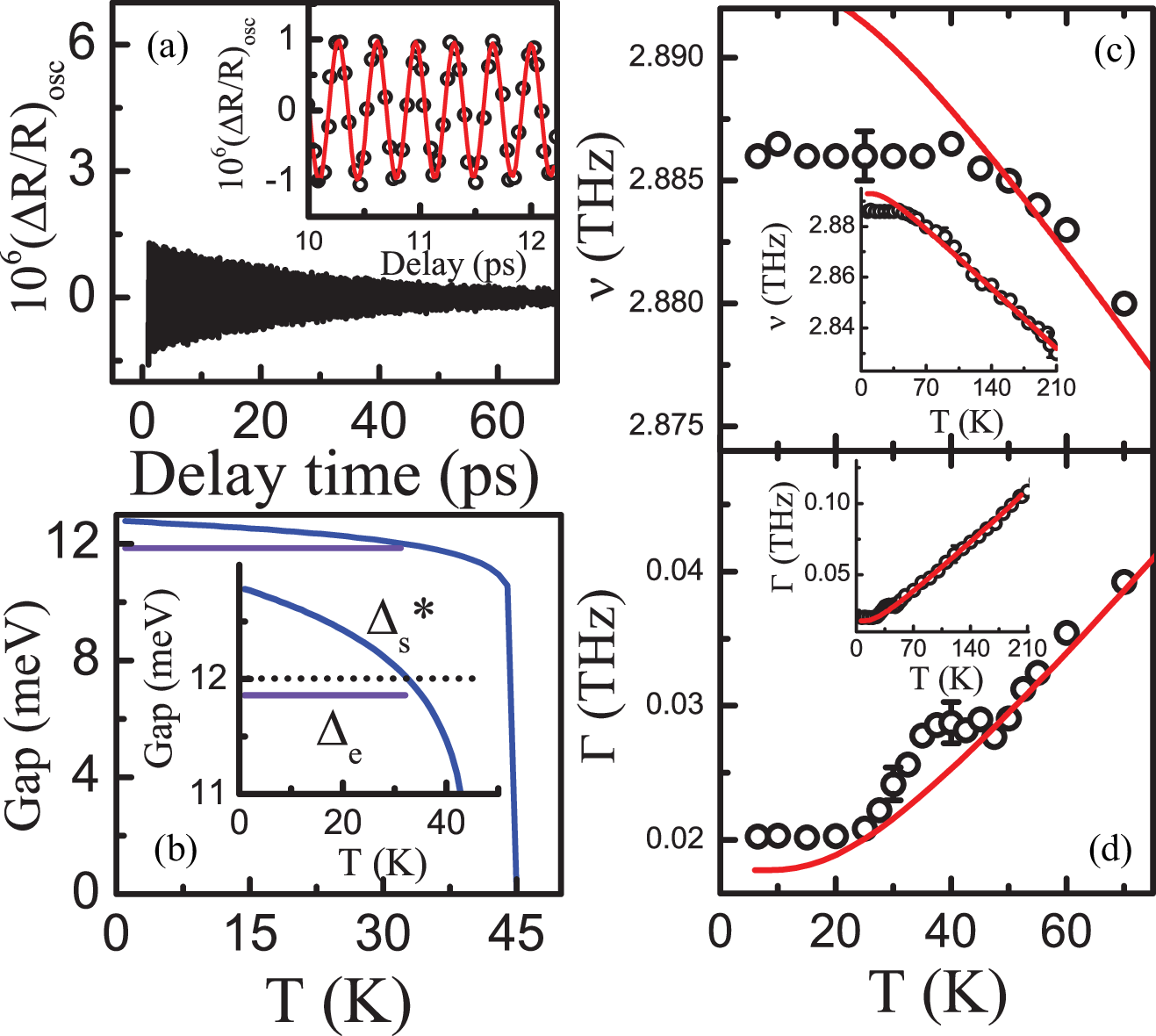}
\caption{\label{fig:CoherentPhonon} (a) Extracted oscillation ($\Delta R/R$)$_{osc}$ and its fit (red line) at 60 K. (b) Temperature evolution of the gaps $\Delta_e$ and $\Delta_s^*$ compared with the phonon energy (dotted line) for $T<T^*$. (c) The $T$-dependence of the oscillation frequency $\nu$ and (d) the damping rate $\Gamma$. The red solid curves are fits to the data using the anharmonic effect model.}
\end{figure}

We can gain more insight on these issues by carefully considering the oscillations in the $\Delta R/R$ signal. It is generally accepted that these terahertz (THz) frequency oscillations, due to coherent optical phonons, are initiated either via the displacive excitation of coherent phonons (DECP) \cite{Zeiger_PRB_1992} or a photoexcitation-induced Raman process \cite{Merlin_SSS_1997} in a strongly absorbing material. More quantitative insight is given by subtracting the non-oscillatory background and Fourier transforming the result, revealing only one frequency component (Figure \ref{fig:CoherentPhonon}(a)). This allows us to fit the oscillatory signal with the expression $(\Delta R/R)_{osc}=Ae^{-\Gamma t}sin(2\pi\nu t+\phi)$, where $\Gamma$ and $\nu$ are the damping rate and frequency, respectively (Fig. \ref{fig:CoherentPhonon}(a)). The $T$-dependence of  $\nu$ and $\Gamma$ is shown in Figure \ref{fig:CoherentPhonon} (c,d). It can clearly be seen that $\nu$ and $\Gamma$ depend almost linearly on temperature above $T^*$ ($\sim 45$ K) but exhibit more complicated behaviour for $T\lesssim T^*$.

The $T$-dependence of the phonon frequency and damping rate is typically explained by the anharmonic effect \cite{Menendez_PRB_1984, Balkanski_PRB_1983, Tang_PRB_1991}. This effect, leading to a renormalization of the phonon energy and damping, usually includes contributions from thermal expansion of the lattice (Gr\"{u}neisen law) and anharmonic phonon-phonon coupling. We can thus model the experimentally measured $\nu$ and $\Gamma$, including these contributions, using \cite{Menendez_PRB_1984, Balkanski_PRB_1983, Tang_PRB_1991}
\begin{align}
\omega(T)=\omega_0+\Delta\omega^{(1)}(T)+A_1[1+2n(\omega_0)], 
\\
\Gamma(T)=A_2[1+2n(\omega_0)],
\label{eq:Phonon-fit}
\end{align} 
where $\omega=2\pi\nu$,  $n(\omega)=[e^{\hbar\omega/k_BT}-1]^{-1}$, and the shift $\Delta\omega^{(1)}$ from thermal expansion is given by $\Delta\omega^{(1)}(T)=\omega_0[e^{-\gamma\int\limits_{0}^{T}(\alpha_c+2\alpha_a)dT^{\prime}}-1]$. Thermal expansion factors $\alpha_i$ $(i=a, c)$ are obtained from Ref. \cite{Henkie_JAlloysComp_1992}. Fig. \ref{fig:CoherentPhonon} (c,d) demonstrates that the above model can explain the $T$-dependent behaviour of $\nu$ and $\Gamma$ above $T^*$. However, it fails to capture the phonon softening in $\nu$ (relative to the model prediction) and the wiggle structure in $\Gamma$, which both appear below $T^*$.  

It is well known that in magnetic materials, phonon renormalization can occur not only through the anharmonic effect, but also through magnon-phonon coupling, which is intrinsically caused by the modulation of the exchange integral $J(r)$ via lattice vibrations \cite{Baltensperger_HPA_1968}. In USb$_2$, the magnetic structure changes from paramagnetic to antiferromagnetic at $T_N$, which could change the magnon-phonon coupling and consequently the phonon softening. However, we did not observe a clear change in $\nu(T)$ at $T_N$ (inset to Fig. \ref{fig:CoherentPhonon}(c)), suggesting that this effect does not play a significant role. 

More insight is obtained by again noting that the large phonon softening occurs below $T^*$, which is the same temperature where the gap $\Delta_s^*$ opens, suggesting that both phenomena are closely related. We first note that $\Delta_s^*$ should be spin-related, since it is derived from $\tau_s$, which diverges when the QP gap opens at $T_N$. This is further supported by the fact that a low energy spin excitation appears below $T^*$ in specific heat measurements (see the supplementary information). These observations indicate that magnons participate in the boson-mediated band renormalization below $T^*$,  leading to the opening of $\Delta_s^*$. This band renormalization increases the effective mass of the bands at the Fermi level with respect to the bare band structure, increasing their DOS. This in turn increases the density of QPs near the Fermi surface, which enhances the screening of atomic forces, leading to the phonon softening seen in our experiments. The observed phonon softening thus provides further evidence for boson (\textit{magnon})-mediated band renormalization, where we note that the details of possible coupling of the spin component to the itinerant electrons remain unknown and need to be investigated further.

At $T^\dagger$ ($\sim 30$ K), the gap $\Delta_e$ opens, which is associated with an additional band renormalization at the Fermi level. As with $\Delta_s^*$, the opening of this gap could influence phonon softening. However, we did not observe any clear change in $\nu(T)$ around $T^\dagger$ (Fig. \ref{fig:CoherentPhonon}(c)). This implies that any enhancement of the DOS associated with $\Delta_e$ is not large enough to modify the phonon frequency, indicating that boson-mediated band renormalization is not the origin of $\Delta_e$. Instead, this gap is likely due to the hybridization of $f$-electron and conduction electron bands, as in other HF materials. This is supported by the fact that $\Delta_e$ is constant below $T^\dagger$, consistent with previous findings in other HFs \cite{Demsar_JPC_2006}. In contrast, $\Delta_s^*$, which deviates from a constant $T$ dependence, has a more complex origin (i.e. boson-mediated band renormalization). More importantly, this reveals that the QP states at/near the Fermi level associated with $\Delta_s^*$ and $\Delta_e$ are quite different.

The damping rate $\Gamma$ of a coherent phonon with energy $E_{ph}$ ($=h\nu$) will be strongly enhanced upon a gap ($\Delta$) opening if $E_{ph}>\Delta$,  due to increased coupling between the phononic and electronic degrees of freedom. In contrast, $\Gamma$ will be unaffected for $E_{ph}<\Delta$ \cite{Zeyher_ZPhyB_1990, Albrecht_PRL_1992}. Here, there are two gaps, $\Delta_s^*$ and $\Delta_e$, which successively open as $T$ decreases. From Fig. \ref{fig:CoherentPhonon}(b), we can see that: (1) for $T<T^\dagger$, $E_{ph}$ is always larger than $\Delta_e$; (2) for $T\gtrsim T^\dagger$, $E_{ph}$ is greater than $\Delta_s^*$. Thus, as shown in Fig. \ref{fig:CoherentPhonon}(d), the damping rate $\Gamma$ below $T^*$ should be always greater than that resulting from only considering the anharmonic effect, since $E_{ph}$ is always greater than one of the two gaps. In addition, since $\Delta_s^*$ gradually increases as $T$ decreases, the coherent phonon energy $E_{ph}<\Delta_s^*$ below $\sim T^\dagger$. Therefore, the damping associated with $\Delta_s^*$ decreases as $T$ decreases from $T^*$ to $T^\dagger$ ($\sim 30$ K). Based on these considerations, the dependence of $\Gamma$ on $T$ is qualitatively expected to show the behaviour in Fig. \ref{fig:CoherentPhonon}(d). Clearly, $T$-dependent phonon damping can intuitively reflect the temperature evolution of the gap structure(s) in strongly correlated systems.

In conclusion, we used ultrafast optical spectroscopy to shed light on the detailed electronic structure of USb$_2$. Temperature-dependent QP relaxation dynamics revealed the opening of three gaps at different temperatures, $T^\dagger$, $T^*$, and $T_N$. The magnitudes of these gaps agree well with previous ARPES results and boson-mediated band renormalization calculations. More insight into their properties is given through the $T$ dependence of the frequency and damping of coherent optical phonons. Strong phonon energy renormalization below $T^*$ also indicates that magnons are the bosons involved in band renormalization, which greatly increases the QP effective mass at the Fermi level. Overall, these findings significantly enhance our understanding of the complex emergent states in USb$_2$, as well as in other $f$-electron systems.

This work was performed under the auspices of the Department of Energy, Office of Basic Energy Sciences, Division of Material Sciences. Los Alamos National Laboratory, an affirmative action equal opportunity employer, is operated by Los Alamos National Security, LLC, for the National Nuclear Security administration of the U.S. Department of Energy under contract no. DE-AC52-06NA25396.


\begin{thebibliography}{text}
\bibitem{Threebands} dHvA and ARPES experiments show that the Fermi surface is composed of two hole-like sheets around zone-center (0, 0), and at least one electron-like sheet located near ($\pi$, $\pi$). Thus, we simplify our calculations by considering the interband scattering processes between three bands.
\end{thebibliography}
\end{document}